\documentclass[aps,preprint,amsmath]{revtex4}
\usepackage{graphicx}
\usepackage[usenames, dvipsnames]{color}
\usepackage{url}
\usepackage[ pdftex, plainpages = false, pdfpagelabels, 
         pdfpagelayout = useoutlines,
         bookmarks,
         bookmarksopen = true,
         bookmarksnumbered = true,
         breaklinks = true,
         linktocpage=all,
         pagebackref=false,
         colorlinks = true,
         linkcolor = BrickRed,
         urlcolor = blue,
         citecolor = BrickRed,
         anchorcolor = green,
         hyperindex = true,
         hyperfigures
         ]{hyperref}

\begin{document}

\title{Military Strategy in a Complex World}

\author{Dominic K. Albino, Katriel Friedman, Yaneer Bar-Yam}
\affiliation{New England Complex Systems Institute}
\author{William G. Glenney IV\footnote{The views expressed in this paper
are those of the authors and do not necessarily reflect the position of
the Department of the Navy, the U. S. Naval War College or the CNO
Strategic Studies Group.}}
\affiliation{Chief of Naval Operations Strategic Studies Group}

\begin{abstract}
A strategy is a plan, method, or series of actions for obtaining a specified goal. A military strategy typically employs the threat or use of military force to advance goals in opposition to an adversary, and is called upon where large scale force is viewed as the way to achieve such goals. 
Strategic thinking is traditionally focused on which part or combination of land, air, and naval forces is most effective. This may be too narrow an approach to accomplishing the ultimate end, which is generally political influence or control---or preventing influence or control by others---and almost never consists of physical destruction itself. In order to broaden the discussion of military strategy, we consider here three distinct effects of inflicting stress on an opponent: a) A fragile system is damaged---possibly catastrophically, b) A robust system is largely unaffected, retaining much or all of its prior strength, c) Some systems actually gain strength, a property which has recently been termed antifragility. Traditional perspectives of military strategy implicitly assume fragility, limiting their validity and resulting in surprise, and assume a specific end state rather than an overall condition of the system as a goal. Robustness and antifragility are relevant both to offense, in attacks against the enemy, and defense, in meeting attacks against one's own forces. While robustness and antifragility are desirable in friendly systems, an enemy possessing these characteristics undermines the premise that an attack will achieve a desired increase in control. Historical and contemporary examples demonstrate the failure of traditional strategies against antifragile enemies---even devastating damage inflicted upon nations or other organizations did not weaken and defeat them, but rather strengthened them, resulting in their victory. Underlying such successful responses are socio-economic or political strengths. Our discussion is a basis for scientific analysis of the historical and contemporary conditions under which distinct types of strategies will be successful and provides guidance to improved strategic thinking.
\end{abstract}

\maketitle

\section{Introduction}
A military strategist is responsible for an overall approach that leads to identifying actions that in turn achieve goals. In these actions, available military power is applied to control an adversary, including its military forces, territory and resources. The strategist assumes that resistance by the adversary will be overcome, leading to behavioral modification in line with the desired outcome. Traditional strategic thinking focuses on how to apply force, including which aspect of the military---land, naval, and air forces---to use in what manner \cite{quade, wayne, bracken, echevarria, johnson, fm1}. Goals may be offensive, defensive or a combination such as an offensive action designed to pre-empt or counter an enemy offensive.

Success of a strategy can often be attributed to the underlying capabilities of a system that is executing the strategy as well as the system that is the target of the strategy. Traditional measures of capability include manpower and firepower, but other capabilities are important as well. These include effectiveness of action (sensitivity to relevant information, effective decision making based upon that information and effective exercise of force), the degree to which the system is damaged by the application of force (robustness), and the ability to recover from damage (resilience). 

\begin{figure}
\includegraphics[width=12cm]{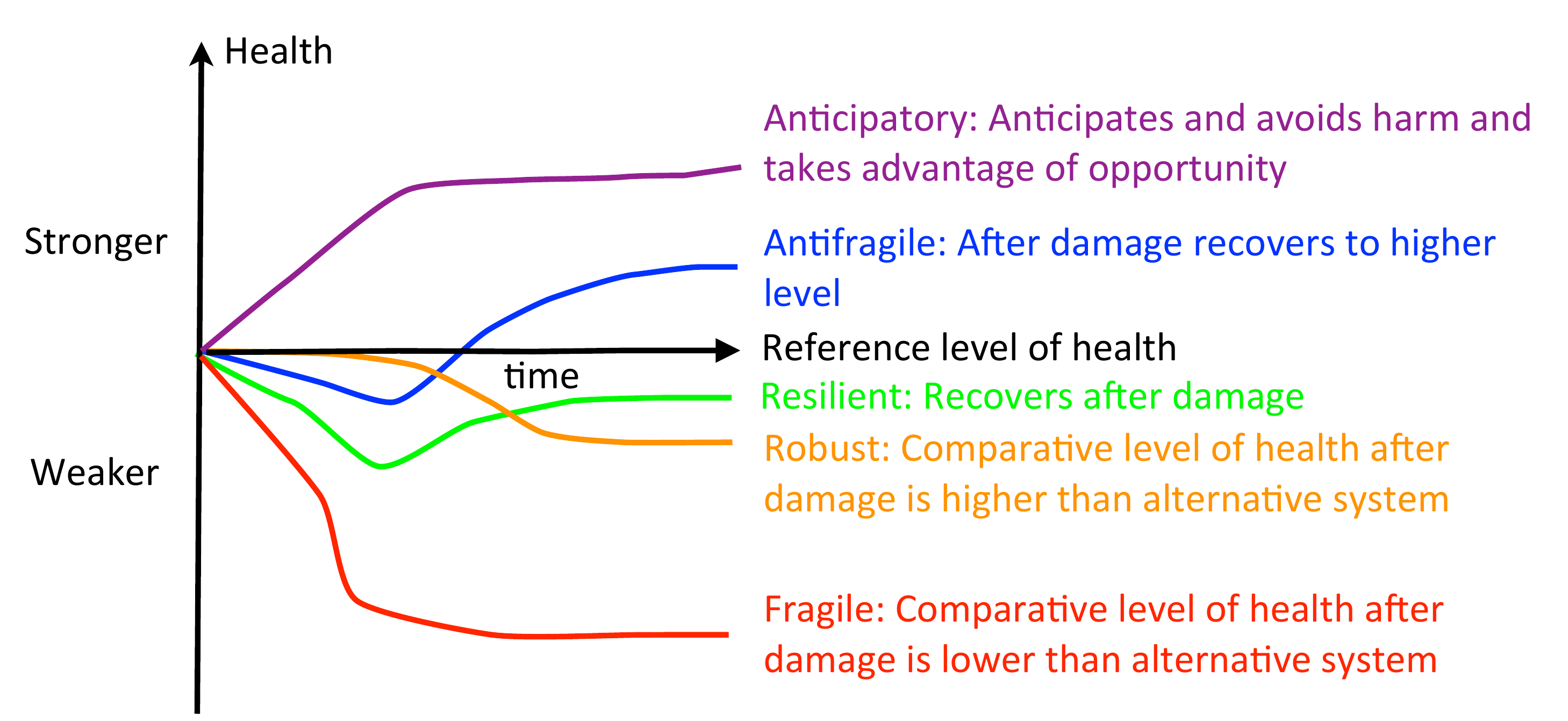}
\caption{Characterization of the changes in system health when subject to external stresses. System properties can include (a) the ability to anticipate, resulting in avoidance of harm or exploitation of opportunities (anticipatory, purple), (b) the ability to recover from damage to a higher level of health than before (antfragile, blue), (c) the ability to recover from damage (resilient, green), (d) the ability to withstand damage (robust, orange), and (e) breaking under stress (fragile, red). \label{fig1}}
\end{figure}

Strategic thinking becomes more complicated when the exercise of force may be ineffective and even counterproductive, leading to defeat rather than victory. The result of exercising force depends greatly on the response to that force over time. We can identify three classes of adversary response that are relevant to the outcomes of conflict: A fragile system is damaged by stress---possibly catastrophically, which is typically the attacker's desired intent and result. A robust system, on the other hand, is largely unaffected by the stress and retains much or all of its prior strength. Finally, some systems actually gain strength from exposure to stress, a property which has recently been termed antifragility \cite{antifragile}. For example, antifragility may occur when an attack motivates increased recruitment that ultimately augments an adversary's military capability. We can combine the response to damage with the ability to anticipate---to act to avoid damage or exploit opportunities---creating a more complete characterization of system properties as shown in Fig. \ref{fig1}.

Traditional ideas about control implicitly or explicitly focus on destruction of enemy forces, the fragility of which is usually an unstated assumption. Some have questioned this approach. Rear Admiral J.C. Wylie asked, ``What kind of control is desired, and under what circumstances will destruction or the threat of destruction bring about the desired measure of control? Judgments of this kind are among the most difficult and speculative of all the problems of strategy'' \cite{wylie}.

In addition to the impact on the adversary, an attack may have deleterious impacts on the attacker's side by depleting military, economic, or political resources, or diverting resources from defense, creating vulnerabilities. It may also establish a persistent disadvantageous pattern of interaction with the adversary or with other groups. Actions involving larger forces can be expected to have disruptive effects on vulnerable socio-economic systems. Even limited force may cause undesired disruption of social order, undesired casualties, and persistent enmity that undermines desired longer term relationships. Finally, attacks may undermine the goals of the attacker by destroying desired resources, or by violating its fundamental values, for example causing civilian casualties. These adverse ``indirect'' outcomes create a paradox for strategists who are focused primarily on direct observable effects. 
It is also important to recognize that a particular goal may or may not be possible given the cultures and social structures of the nations or groups involved. 

The variety of ways initial success of military force can be undermined are manifest in examples ranging from ancient conflicts to modern wars. From the Punic wars to recent conflicts such as the Iraq War, as well as the classic examples of Napoleonic and German invasions of Russia in 1812 and 1939, many examples show that initial success in defeating enemy forces, or success in more limited measures such as occupying territory, does not necessarily translate into ultimate success in achieving desired goals \cite{johnson, gordon, kagan, lazenby}. Even in the case of a successful defeat of a fragile enemy, the actual outcomes may not be the desired outcomes. 
A classic example is the Allies' WWI victory and the subsequent political and economic sanctions imposed on Germany. The sanctions are generally considered to have contributed to the conditions that resulted in WWII \cite{paxton, grunberger}. Another example of inherently ineffective attacks is in deterrence by mutually assured destruction \cite{schelling1966arms}, a strategy adopted in the Cold War confrontation of the US and USSR. In this context any attack would result in such a high level of mutual destruction that no rational objective could be served. The strategy was conflict deterrence through the threat of attack.

The complications of strategy arise because the exercise of force is a means to accomplish other objectives rather than a goal in itself, the goals of the strategy are often too narrowly defined as a specific end state, while the actual objectives are ongoing societal conditions. In the early 1800s, the military theorist Clausewitz stated that the goal of war is ultimately political \cite{onwar}. While destruction is sometimes envisaged, modern conflict almost always occurs in the context of ongoing, mutually beneficial economic, social, and political relationships. To be successful, strategists must account not only for short term effects, but also impacts on future conditions. Military strategies should therefore consider a variety of psychological, economic, and political measures along with, or instead of, targeted military force. Even when advisable, military action may be one step, but is not the only step, in a process of achieving a desired relationship between two or more peoples, or establishing a desired condition within a system. 

The analysis of strategy through game theory is often focused on games with well known payoffs. However, a key question in real word strategies is determining the mapping of real world options onto payoffs. In multi-move games, successful strategies need not be the ones that make the first destructive strike ``successful'' or give rise to early advantages in naive measures of success. This is the case whether or not payoffs are clear, and even applies in well-defined conflict abstraction games such as chess. Determining when and how to apply force to achieve desired objectives is essential. Our analysis points to the existence of generic conditions in which force is counter-indicated even though it may appear to be the direct path to success. Effective means are needed for evaluating both the consequences of direct application of force and of the alternatives to force. The concepts of fragility, robustness and antifragility are useful when the characteristics of the response are similar across a range of possible acts; they are attributes of the system and not of the particular action. 

This paper marshals theoretical tools to suggest the importance of indirect effects, both at the level of selecting the type of action, and at the more focused level of pursuing specific military objectives. We describe the ramifications of complexity \cite{dynamics} for strategic military thinking. We give special consideration to military forces encountering non-fragile adversaries. Learning and adaptation enable antifragility, violating traditional strategic assumptions. 
We also give preliminary indications of how strategists can apply these concepts to better assess risks posed by adaptable enemies and enable their own forces to take advantage of opportunities to reverse apparent setbacks.

This work contributes to a scientific literature that expands the understanding of the origins and consequences of violent conflict \cite{lim, rutherford, gros, friedman, lagi1, lagi, south, gard-murray}. Previous efforts in this direction have analyzed the conditions under which ethnic violence \cite{lim, rutherford, gros, friedman} and food-insecurity based riots and revolutions \cite{lagi1, lagi, south} arise. Recent analysis of whether and when violent revolutions achieve their objectives \cite{gard-murray} is directly related to this work in that it analyzes the effectiveness of large scale violence in achieving strategic goals. The role of complexity and scale in military forces and success has also been discussed \cite{making,civilization, military}.

\section{What is strategy?}

A strategy identifies a method or set of methods to obtain a particular end. Traditional and modern strategic thought, such as that codified by the U.S. Army War College model, frames military strategy as a static plan with narrow, clearly defined results made in advance of conflict. Strategy, in this perspective, exists at a higher level of planning than the operational and tactical decisions made during the course of a conflict. Classical and modern strategists have almost exclusively treated armed forces as fragile systems: The enemy's military capacity will collapse when subjected to sufficient pressure that is correctly applied. Each strategist aims to reduce the opponent's ability or willingness to fight while maintaining his own. U.S. military doctrine describes two ways of accomplishing this goal: 
annihilation, which seeks to totally incapacitate the enemy, and erosion, which attempts to impose high costs on the opponent. 
In general, the foundation of strategy is determining a method to cause the collapse of the enemy's organization due to hopelessness or confusion. At most, traditional strategists seek to make their forces as robust as possible, and in all cases comparatively more robust than the enemy.  
More indirect strategic thinkers such as Sun Tzu, Andre Beaufre, and John Boyd have held up the ideal of winning a psychological victory while avoiding hugely destructive engagements \cite{art, beaufre, boyd}. Annihilationists like Clausewitz scoffed at the feasibility of this notion and insisted the way to win was to cause the enemy to collapse by destroying his center of gravity \cite{onwar}. Taking advantage of the enemy's vulnerability to damage and consequent loss of capability is a common thread across traditional schools of thought. This results from a fundamental assumption that military forces are fragile and degrade under stress, so that any stress applied brings the force closer to destruction or defeat.

\section{Confounding Forces in Strategic Theory}

The history of military encounters, as well as complex systems theoretical insights, suggest that traditional military strategy is too limited an approach to reliably achieve desired goals. Defeat, or outcomes far from those commonly envisaged, can arise from a variety of scenarios when military force is applied. These confounding effects sometimes undermine the premise that direct military force should be the basis of strategy, but they more generally should be considered an important aspect of military strategic thinking. Recognizing the potential limitations of applied force can improve both proactive solutions and countermeasures to an adversary's action.

Traditional military strategy might be considered applicable when the goal is to substantially destroy, dominate or subjugate in a policy of expansionism, or countering an adversary with such a goal. Under these conditions, complications seem minimal. However, even in such comparatively simple circumstances, there are complications that affect both offensive and defensive approaches.

As described in the Introduction, a wide variety of examples, ranging from ancient conflicts including the Punic wars, to recent conflicts such as the War in Iraq, show that initial success in defeating enemy forces does not necessarily translate into ultimate strategic success \cite{johnson, gordon, kagan, lazenby}. 
In addition to direct failure to defeat an enemy with the application of force, there are a variety of reasons for a strategy to be confounded or non-optimal including: 

\emph{Larger and other goals}---Objectives forming the basis of a military strategy always exist in a web of other explicit or implicit goals that complicate the development and directly impact the environment within which military strategy is executed. These goals can include political, social, economic and environmental goals as well as humanitarian and other ethical constraints \cite{smith}. Moreover, objectives and goals may not be static, shifting in main or in detail, necessitating a corresponding shift in strategy. A successful strategy is therefore one that recognizes not only the immediate objective, but also the existence of other goals that constrain methods or enable multiple goals to be achieved.

\emph{Conflict and cooperation}---Conflict arises when the objectives of different parties are seen as incompatible. Nevertheless, actors can often choose among multiple objectives and associated strategies, some of which are contradictory and others of which are not. For example, an objective of national economic success may be alternatively framed in terms of conquest of land and other resources, or mutual advancement through economic cooperation and competition. Much of the change in global strategies of war and peace in the past 100 years has been achieved through a shift from the former to the latter approach. Given the destructive nature of large-scale conflict, cooperative strategies often have comparative advantage due to the risks and costs of conflict. Inflexible criteria for judging outcomes may overlook such beneficial opportunities. Thus, the effective use of military strategy is more limited when a decision to employ destructive military force precedes an assessment of what such force can accomplish. A more effective approach compares outcomes among a broader range of alternatives. When actors are unwilling or unable to change their objectives, conflict becomes unavoidable. Today, a key challenge is identifying how to transition to new forms of cooperation in the face of a diversity of local values and objectives \cite{power}. 

\emph{Indirect effects as disadvantage}---Indirect effects impact stated objectives as well as larger explicit or implicit goals. Awareness and realistic evaluation of these indirect effects are needed for effective strategic planning, including difficult to consider, longer term socio-economic impacts and psychological responses, as well as impacts on the environment and context in which a conflict occurs.  
An initial defeat may trigger mobilization of the larger society, propelling an adversary's military force to subsequent victory. This suggests that military strategy must consider an evaluation of societal strength and weakness, rather than just the existing military forces, as a longer term indicator of success. Indeed, economic strength, social resolve, and cultural support---or an undercutting lack thereof---have substantial explanatory power from the Napoleonic wars \cite{goodlad} to WWII \cite{riefler} to the Cold War \cite{gelb} and modern hotspots around the globe \cite{nye}. 

Why, then, are military forces not commensurate with societal strengths? In brief, we can hypothesize that aggressive antagonists prioritize and prepare for large scale military action, while opponents may not begin substantial preparation until attacked. Indeed, those who aspire to power may pursue militarism as a means of advancement precisely because their societies are otherwise weak. In contrast, by focusing on economic and other activities, peaceful countries may develop reserves of strength not measured in military terms. This results in military capabilities not consistently proportionate to societal strength. Antagonistic interactions between countries may expose underlying strengths. By creating existential threats, aggressors may motivate those who are attacked to use extraordinary means to enhance their own strengths. Once the societies respond, they may well be significantly more capable and therefore more successful militarily. 

These widely observed indirect effects can be extended to include a variety of indirect effects that have increasing importance in the modern world. As the example of the WWI defeat of Germany illustrates, socio-economic disruptions due to conflict can breed their own destructive consequences with broader implications. More generally, indirect effects of military force should include the local disruption of social stability and related adverse socio-economic conditions. Such destruction and the absence of local order enable the growth of criminal and terrorist networks having substantial local power and global reach. This adverse effect is distinct from the strengthening of adversaries, as the resulting organizations need not be directly associated with the original conflict or adversary but instead arise in the fertile context of the prevailing socio-economic disorder. 

That traditional military strategy focuses on the effect of military action on military forces and not on the larger societal context \cite{echevarria, johnson, smith} is reflected in the writing by US Army Chief of Staff Raymond Odierno, ``This capstone doctrine publication frames how we think about the strategic environment, develop and refine doctrine,\ldots Warfighting is our primary mission. Everything that we do should be grounded in this fundamental principle. \ldots Nobody in or outside the military profession should mistake the Army for anything other than a force organized, equipped, and trained for winning the Nation's wars.'' \cite{fm1}. 

Despite a primary focus on combat operations, some strategic military thinking has begun to turn to surrounding, contextual problems. For example, Lt. General William Caldwell IV writes, ``The greatest threat to our national security comes not in the form of terrorism or ambitious powers, but from fragile states either unable or unwilling to provide for the most basic needs of their people\ldots  Military success alone will not be sufficient to prevail in this environment. To confront the challenges before us, we must strengthen the capacity of the other elements of national power,\ldots At the heart of this effort is a comprehensive approach to stability operations that integrates the tools of statecraft with our military forces, international partners, humanitarian organizations, and the private sector.'' \cite{fm3} 

\emph{Indirect effects as advantage}---A strategy often assumes a fixed set of conditions or constraints imposed by existing natural and social context. But, if one considers the conditions malleable, the indirect effects of changing the conditions may improve the set of options available. The environmental constraints may be physical or social. An important source of environmental change is changing the roles of third party actors within the environment, especially those with conflicting objectives. Likewise, the strategist is part of the environment other actors reckon with, and may change the context in ways to support or obstruct other actors' objectives. A common example in recent years is the development of improved relationships with local populations as a means to strengthen one's capabilities and weaken an adversary. This situation arises for a force operating in hostile territory with the objective of defeating part of a global terrorist network. Actions to improve relationships with a portion of the local population enable access to local information networks relevant to tracking adversary activity and reducing enemy recruitment. A very different modification is a scorched earth strategy---the destruction of agriculture, transportation, communications, 
and industrial resources that may help enemy forces. The approach, used by Sherman in the Civil War, 
and Russia against the Germans in WWII, may provide short term indirect effect advantage but is forbidden under international law \cite{geneva} because of longer term adverse consequences. (As discussed below, the former approach assumes a significant degree of resilience or even antifragility, while the latter approach assumes fragility.)

\emph{Uncertainty and chaos in conflict}---While widely recognized, it is important to account for unpredictability undermining strategic planning. Uncertainty arises from the absence of crucial information as well as the unpredictability of events as they unfold. Conflict is an engagement between groups that are actively and purposefully interfering with each other's plans. The development and implementation of a strategy by one side, counter strategy by another, a new strategy, a new counter strategy, and so forth \cite{schelling} means that available information is often insufficient to predict outcomes. When the capabilities of opposing sides are comparable, the outcome cannot be certain prior to the tactical engagement. Even in conflict where victory of one side appears assured, the details, process, and side-effects of such a victory are often unknowable beforehand. Moreover, in processes of action, reaction, and interaction, even small deviations may cause rapidly enlarging effects that undermine predictability \cite{kellert, marion}. Reliance on forecasts is likely to result in misleading information and misplaced confidence \cite{lorenz, silver, ormerod}. Effectiveness of pre-specified sequential strategies is limited because they rely on anticipation of intermediate outcomes. Chaotic unpredictability is embodied in the adage ``no plan survives first contact with the enemy.'' Effective strategies in such conditions focus on enabling responses to a set of possibilities rather than specific events. 

\section{System Response to Attack}

An essential aspect of strategy is understanding the effect of actions on future capabilities. The alternatives of fragility, robustness, resilience, antifragility \cite{antifragile} and anticipatory response \cite{baryamepstein} must be considered. 

Fragility is characterized by deterioration under stress. This is the generally expected result of military attack, and the intent of strategy to degrade an enemy's forces to yield a desirable control. Attacked by a large enough force, and over a sufficient period of time, any system is fragile.
The expectation that large military forces are needed to defeat an enemy arises from the common assumption that increasing amounts of stress result in greater than linear increase in damage to fragile systems \cite{pp}. The damage from 10 shocks is generally much less than the damage from a single shock with 10-fold force. This is consistent with standard military strategy that strives to deliver large shocks to an enemy that is physically or psychologically fragile. Conversely, where possible, the maximum size of shocks that a fragile friendly system suffers should be reduced in favor of multiple smaller size ones. Nevertheless, a very large number of small scale shocks, as in attrition warfare, can also be employed to devastating effect.

Robustness and resilience characterize systems that tolerate stress. This may be due to relative scale, such as when a person is hit by a ping pong ball. 
Robustness and resillience may also result from the ability of the system to recover from a shock, a blade of grass is affected, but need not be damaged, by either a gentle breeze or a storm. Robustness and resilience are {\em de facto} invulnerability to a specific stress, rendering such specific attacks futile. If a friendly system possesses this property, it can devote resources to other areas of importance. Attacks to which an enemy is robust or resilient are a wasted effort. A historical example of mutual invulnerability is the stalemate that persisted between the Colombian military and FARC and ELN guerrilla forces for decades \cite{farc}. 

``Antifragile'' is used to mean the direct opposite of ``fragile''---a fragile system is one that deteriorates when stressed, an antifragile system grows stronger. 
The term ``antifragility'' was coined by Nassim Taleb because the concepts of robustness and resilience commonly contrasted with fragility do not indicate a truly opposite meaning \cite{antifragile}. Resilience and robustness do not imply gains in strength, as does antifragility.

Antifragility turns setbacks into advantage in the course of an ongoing conflict assuming conflict is not resolved by a single strike. When one of the participants in a conflict is antifragile it confounds opponents' traditional plans by strengthening, rather than weakening, under stress. When one's own (friendly) system is antifragile, stress and conflict imposed by the adversary should be sought to improve the function of the system. When an enemy system is antifragile, stress and conflict should be avoided in order to limit the improvement of that system. 
An antifragile system must be opposed by means of other dimensions of activity, for example responding to a militarily antifragile system through political, social, or economic fragility \cite{marsden, peters}.

One archetype of an antifragile force is an insurrection that gains members and support as a result of the conditions of conflict. Contemporary examples include the Taliban from the time of the Afghan-Soviet war to the present day in Afghanistan and Pakistan \cite{tanner}, as well as Al-Qaeda throughout the Islamic world \cite{al-qaeda, mccormick}. Historically, antifragility is a common thread in wars of national independence, including the American Revolution \cite{miller}. In these cases, actions by the occupier fuel their opponents' will, unity, and organization. 

A second archetype is a state subjected to strong, unexpected attack which, as a result, mobilizes on a vaster scale than would otherwise have occurred, thereby causing the defeat of the attacker. Examples of this type include Rome during the Second Punic War \cite{britannica}, Soviet forces in response to German Operation Barbarossa \cite{glantz}, and the United States in the wake of Pearl Harbor \cite{ambrose}.

A state or other group has a strong interest in formulating a strategy that promotes and exploits its own antifragility. These states should also understand when adversaries or potential adversaries are themselves antifragile, so that strategic risk assessments can be accurate, and strategic plans are effective. Where difficult to establish, estimates of enemy forces' antifragility have critical strategic implications.

Since the concept of antifragile systems is generally counterintuitive, we will consider how such capabilities arise and how to foster them in the subsequent sections of this paper.  

For completeness, we note, however, that there is an alternative approach to consider for the response of a system to stress: anticipatory response, which enables response before the stress occurs.
Anticipatory response is a key capability of complex systems \cite{baryamepstein} and is distinct from both robustness and antifragility \cite{talebfootnote1}. Anticipation involves sensitivity to signals and the ability to interpret them as indicators of potential future harm (or benefit). Anticipation also requires dynamic response that prevents harm or takes advantage of opportunities. The difference between robustness and anticipatory response can be seen in a biological context in the generic response capabilities of plants and animals. While plants survive when parts of them are eaten by animals, animals often survive by fighting or fleeing. This capability reflects sensor-action coupling through an intermediate processing system in response to information in  the signal about future events. Since the signal for the impending harm has a smaller impact than the harm itself, this is the origin of the term ``sensitivity'' in the ability to achieve anticipatory response. Such sensitivity enhances complex systems capabilities across a variety of contexts. 

\section{Sources of Antifragility: Evolution, Learning, and Adaptation}

The foundations of antifragility as a basis for military strategy are evolution, learning, and adaptation. An antifragile system must be dynamic, and can, at the level of human institutions, be intentionally constructed exploiting mechanisms that give rise to antifragility. 

Evolution is a repeated process of selection and reproduction with variation: Varieties that outperform their predecessors become the basis for the next round of variation, competition and selection. Antifragility emerges in a system when its components are selected and replicated according to their contribution to system effectiveness. As a result, the entire system is strengthened when shocks lead to selection of effective components. For evolution to take place, there must be a source of variation, a selection process to weed out less successful variants, and a means for successful variants to replicate. In biology, variation comes from heritable random mutation and recombination of the genome. Selection arises through the competition for survival and reproduction. In the context of human society, variation often arises from creativity and innovation in ideas or behaviors. Selection and replication occur when successes and failures lead others to make changes in their behavior by adopting the innovations in ideas and behaviors. Stresses cause selection pressure that eliminates less successful variants of behavior. When the adoption of more successful behaviors occurs, the entire population can better withstand similar stresses and thus the system as a whole is strengthened. Evolution contributes to antifragility at the collective or organizational level, but individual components remain fragile due to the process of selection. Over time, selection may result in individuals that possess antifragile individual-level characteristics, such as rapid learning or adaptation.

Learning occurs when exposure to stress provides information about the challenges faced. An individual gains antifragility when he or she (a) has the opportunity to experiment with different responses and (b) makes use of accumulated knowledge about how to successfully respond to a stress. Information about a new stress can be used to repeat successful strategies or avoid unsuccessful ones. Learning can also be considered to be the evolution of ideas, where ideas that are evolving determine which responses are chosen. Learning can occur through multiple applications of the same information in simulated trials of variant ideas. War gaming is one common mechanism that the military can use to develop and evolve strategies. 

Adaptation is also a source of antifragility. Adaptation allows for an increase in the effectiveness of a response to a particular stress after that stress has manifested itself. Human adaptation takes many forms, some of which occur without conscious action. An example is human muscle growth and increase in cardiovascular capacity. When stress is placed on the muscular or cardiovascular systems, such as during athletic training, these systems respond by increasing capacity, resulting in a more capable system. This genetically-programmed adaptive response enables the body to better match abilities to environmental demands. Other forms of adaptation require conscious reaction, such as observation and response to ongoing events. Conscious reaction can involve either anticipatory or reactive response in the context of damage. Anticipatory response, described above, is a distinct mode of effective response. Reactive response is a form of antifragility---when damage occurs, the system reevaluates its behavioral options and chooses a more effective one that will enable better consequences in the future. 

The relevance of learning and adaptation to antifragility makes clear that antifragility is already central to military doctrine in training at an individual and tactical level. Military training is focused on the experience of adverse and challenging conditions. Troops are frequently exposed to a variety of combat simulations and drills, including large scale war games. Setbacks are often a source of inspiration and learning for subsequent actions. Yet, the concepts of evolution, learning and adaptation are infrequently extended to military strategic planning to enable better response to the dynamic and unpredictable nature of ongoing conflict. 

\section{Social (Anti)fragility in the Dynamic Response to Conflict}

In order to discuss antifragility more concretely for a military context, we consider a scenario that is relevant to historical cases in which a strike by one combatant disables the second. However, the second combatant is able to become stronger than it was before, and subsequently defeats the first combatant. This scenario was relevant to Rome during the Second Punic War \cite{britannica}, Soviet forces after German Operation Barbarossa \cite{glantz}, and the United States in the wake of Pearl Harbor \cite{ambrose}. The general discussion is also relevant  to many contemporary examples from the Taliban in the Afghan-Soviet war to the present day conditions affecting the US in Afghanistan, Pakistan, Iraq and Syria \cite{tanner}, including Al-Qaeda and ISIS \cite{al-qaeda, mccormick}.

In a scenario where a strategically antifragile relationship exists, an initial attack disables the mobilized forces of a country. The subsequent response requires: 1) Resources that survive an attack, including a previously unmobilized population and economic resources, and 2) Values that  reject the terms of defeat, and encourage militarization. The role of values in conflict reflect culture, ideology, religion and nationalism. Traditional strategic analysis hinges on a loss of will as key to collapse. The collapse of will is assumed to occur in the wake of destruction and by values accepting the conditions of defeat and the imposed control defeat implies. The role of accepting or not accepting defeat is apparent in historical examples we have cited for antifragility including Russia's response to Napoleon's attack and Iraq's response to the US in the Iraq war. 

A traditional analysis of military capabilities focuses on mobilized forces and does not account for the unmobilized forces that may be mobilized in response to an attack. It is apparent from our discussion that strategic analysis should include the potential unmobilized forces in conjunction with values that would lead to their mobilization. A  recognized failure of such analysis occurred in the invasion of Iraq, where stated expectations suggested that much of the populace would welcome an invasion that deposed Saddam Hussein \cite{welcome}. However, such a response did not occur \cite{nowelcome}.   

\section{Military (Anti)fragility in the Dynamic Response to Conflict}

While socio-economic resources and popular resistance to imposed control are a key to understanding systemic antifragility, there are other properties that enable shorter time scale dynamic adaptation and antifragility in the context of a conflict.
 
One general organizational lesson from evolution is that to make an organization antifragile, components of the organization should have a diversity of behaviors rather than uniformity through standardization. This diversity enables the recognition of variants that are more or less successful in response to a stress. The better performing variants can be replicated to improve the response to future stresses. In order for this process to take place, measures of success and failure must be incorporated in feedback processes that promote learning across the organizational units. Learning facilitates adaptation that is important both tactically and strategically. The incorporation of intentional diversity into unit behavior goes against the military tendency toward homogenization and ``harmonization.'' Such homogenization is consistent with and motivated by considerations of large scale efficiency but limits complexity and adaptability \cite{dynamics}. Homogenization is also appropriate when there is predictability about what actions are needed. Homogeneity is ineffective when there is a lack of predictability about the types of stress to be faced. Complexity in the conditions and needed response to them should be met by increasing variety rather than homogeneity.

In general, the complexity of conflicts implies that there is a great deal of information commanders may be missing about what will be effective. New information is important, and continues to be vital throughout a conflict because of the declining usefulness of older information in a volatile and transient context. As a consequence, there are compelling reasons to compress the timeframe and reduce the cost of learning; in other words, ``fail early, often, and small.''

To evaluate the antifragility of a force in a given setting, one should consider: 
the cost of gaining relevant information (are major engagements the only way of learning about the enemy's intentions and limitations?); the time frame of gaining and incorporating information into tactics and strategy (how long does it take to analyze an engagement and adapt doctrine?); the ``perishability'' of information (how rapidly does information become obsolete?); the importance of new information (how much does new information impact on the outcomes of strategic or tactical actions?); and the evolvability of the system (is it evolving fast or slow?).

Strategists should consider both their learning rate and that of the adversary. They should ask questions like: If the enemy is antifragile because different parts of his organization are practicing different approaches, can we get him to standardize? Can we convince the enemy not to trust mid- and low-level decision-makers and calcify the command structure into a more rigid hierarchy by strengthening the top leadership? Are there opportunities to enhance the adaptation rates of friendly forces? Are there ways to inhibit enemy force adaptation?

Reducing the enemy's capabilities while maintaining one's own \cite{fm1} may involve surprising actions in the context of antifragility. In particular, attacking leaders may actually be counter to preventing antifragility. Strengthening central leadership generally reduces antifragility of enemy forces. 

Among the origins of fragility in conflict are: 1) Reliance on a military force with specific manpower and materiel, 2) Reliance on controlling the initiative and an advantageous strategic and tactical position, 3) Limited logistics and resource availability, 4) Popular disaffection from existing governance structures, 5) Minimal differences between governance structures and value systems of opponents, and 6) Psychological vulnerability due to defeatism.

In addition to mitigating the above six sources of fragility, to make a force more antifragile, one should work to strengthen the society through its internal cohesion, organization, and goals, as well as its economic structures. In the context of military operations, one should develop mechanisms that shorten the time and cost of learning. To make an enemy force fragile, lengthen the time frame and increase costs for learning, and provide terms of defeat that are consistent with public values.

\section{Conclusion}

A strategy plays out in a context of interactions between actors with different objectives and capabilities. These objectives and capabilities and the environment in which they exist are rarely if ever static. An approach to strategy that accounts for response capabilities, including fragility, robustness, resilliance, antifragility, and anticipatory response can anticipate better the outcomes of planned action. This involves identifying relative strengthening or weakening of capabilities of actors as a result of their interactions over extended conflicts. An interaction that leads to a short term advantage assuming fragility may not lead to a long term advantage in the face of antifragility. An effective strategy should enhance its own force's ability to adapt in the face of setbacks to produce long term gains. Strategists should consider military conflict one process in an on-going relationship between groups, with the indirect effects on the political, social, and economic aspects of that relationship as, or more important than, the direct physical effects on the enemy forces.

\end{document}